\documentclass[preprint,pre,aps,superscriptaddress,showpacs]{revtex4}

\usepackage{amsmath}
\usepackage{graphicx}

\begin{document}

\title{Steady state properties of a mean field model of 
driven inelastic mixtures.}

\author{Umberto Marini Bettolo Marconi}
\affiliation{
Dipartimento di Matematica e Fisica \\
Universit\'a di Camerino, 62032 Camerino, Italy}
\affiliation{
Istituto Nazionale di Fisica della Materia, Unit\`a di Camerino, Camerino, Italy}

\author{Andrea Puglisi} 
\affiliation{
Dipartimento di Fisica \\
Universit\'a ``La Sapienza'',P.le A. Moro 2, 00198 Roma, Italy}
\affiliation{
Istituto Nazionale di Fisica della Materia, Unit\`a di Roma, Roma, Italy}

\date{\today}

\begin{abstract}
We investigate a Maxwell model of inelastic granular mixture under the
influence of a stochastic driving and obtain its steady state
properties in the context of classical kinetic theory.  The model is
studied analytically by computing the moments up to the eighth order
and approximating the distributions by means of a Sonine polynomial
expansion method.  The main findings concern the existence of two
different granular temperatures, one for each species, and the
characterization of the distribution functions, whose tails are in
general more populated than those of an elastic system.  These
analytical results are tested against Monte Carlo numerical
simulations of the model and are in general in good agreement.  The
simulations, however, reveal the presence of pronounced non-gaussian
tails in the case of an infinite temperature bath, which are not well
reproduced by the Sonine method.
\end{abstract}

       
\maketitle

%

\section{Introduction}

Granular materials, a term coined to classify assemblies of
macroscopic dissipative objects, are ubiquitous in nature and play a
major role in many industrial and technological processes.

Interestingly, when a rarefied granular system is vibrated some of its
properties are similar to those of molecular fluids, while others are
unique and have no counterpart in ordinary fluids \cite{general}. A
spectacular manifestation of this difference can be observed in a
driven mixture of granular particles: if one measures the average
kinetic energy per particle, proportional to the so called granular
temperature, one finds the surprising result that each species reaches
a different value. Such a feature, observed in a recent experiment
\cite{Menon}, is in sharp contrast with the experience with other
states of matter. At a more fundamental level such a behavior is in
conflict with the Zeroth Law of thermodynamics.  This principle states
that when two systems globally isolated are brought into thermal
contact they exchange energy until they reach a stationary state of
mutual equilibrium, characterized by the same value of their
temperatures.  A corollary to such a principle is the statement that
the thermal equilibrium between systems A and B, i.e. $T_A=T_B$, and
between A and C ($T_A=T_C$) implies the thermal equilibrium between B
and C \cite{Pippard}.

On the contrary, when two granular system A and B subject to an
external driving force exchange energy they may reach in general a
mutual equilibrium, characterized by two constant but different
granular temperatures $T_A$ and $T_B$.  In addition even when A and C
are in equilibrium at the same temperature $T_C$ and B and C also have
the temperature $T_C$ one cannot conclude that A and B would be in
mutual equilibrium at the same temperature.  In other words, one of
the most useful properties of the temperature, i.e. the independence
from the thermal substance is lost when one deals with granular
materials.

In the present paper we shall investigate the properties of a simple
model of two component granular mixture. The motivation of our study
relies in the fact that granular materials being mesoscopic objects
are often constituted by assemblies of grains of different sizes
and/or different physical and mechanical properties. The study of
granular mixtures has attracted so far the attention both of
theoreticians \cite{Dufty} and of experimentalists \cite{Menon}. In
particular Garz\'o and Dufty have studied the evolution of a mixture
of inelastic hard spheres in the absence of external driving forces, a
process termed free cooling because associated with a decrease of the
average kinetic energy of the system, i.e. of its granular
temperature.  During the cooling of a mixture, which can be
homogeneous or not, according to the presence of spatial density and
velocity gradients, each species may have different granular
temperatures, although these may result asymptotically proportional,
i.e. they might decrease at the same rate.

Menon and Feitosa instead studied several mixtures of vibrated
inelastic grains and reported their failure to reach the same granular
temperature.

In the present paper our interest will be concentrated on the
statistically stationary state obtained by applying an energy feeding
mechanism represented by a stochastic driving force.

The most widely used model of granular materials is, perhaps, the
inelastic hard sphere model (IHS) \cite{IHS}, which assumes the grains
to be rigid and the collisions to be binary, instantaneous and
momentum conserving. The dissipative nature of the collisions is
accounted for by values less than $1$ of the so called restitution
coefficient $r$.  Even such an idealized model represents a hard
problem to the theorist and one has to rely on numerical methods,
namely Molecular Dynamics or Event Driven Simulation, or to resort to
suitable truncation schemes of the hierarchy of equations for the
distribution functions. One of these schemes is represented by the
Boltzmann equation based on the molecular chaos hypothesis, which
allows to study the evolution of the one-particle distribution
function. Its generalization to the two component mixture of inelastic
hard spheres has been recently considered by Garz\'o and Dufty.

Our treatment will depart from previous studies, because we have
chosen an even simpler approach based on the so called gases of
inelastic (pseudo)-Maxwell molecules. These gases are natural
extensions to the inelastic case of the models of Maxwell
molecules~\cite{Ernst1}, where the collision rate is independent of
the relative velocity of the particles. Such a feature greatly
simplifies both the analytical structure of the Boltzmann equation and
the numerical implementation of the algorithm simulating the gas
dynamics. Although the constant collision rate is somehow unrealistic
one may hope to be able to capture some salient features of granular
mixtures, and in particular to reach a better understanding of their
global behavior, because the model lends itself to analytical studies.

This type of approach to granular gases had recently a surge of
activity since the work of Ben-Naim and Krapivsky~\cite{BenNaim,Krapivsky}, our
group~\cite{Balda1,Balda2,Umbm}, Ernst and coworkers~\cite{Ernst2} and
Cercignani and collaborators~\cite{Cercignani,Carrillo}, and is
providing a series of new important results concerning the energy
behavior and the anomalous velocity statistics of granular systems.

The paper is organized as follows: in Section 2 we define the model
and deduce the associated Boltzmann equations governing the evolution
of the velocity distribution functions of the two species and set up
their moment expansion; in section 3 we shall the exact values of the
stationary granular temperatures; in section 4 we consider the moments
up to the eighth order and compute the distribution functions by means
of the so called Sonine expansion. In section 5 we simulate on a
computer the dynamics of the inelastic mixture and compute numerically
the distribution functions and compare these with the theoretical
predictions. Finally in 6 we present our conclusions.
 

\section{Definition of the model}

In the following we shall consider a mixture of Maxwell inelastic
molecules subject to a random external driving.  Let us consider an
assembly of $N_1$ particles of species $1$ and $N_2$ particles of
species $2$. For the sake of simplicity we assume the velocities
$v_i^{\alpha}$, with $\alpha=1,2$, to be scalar quantities. The two
species may have different masses, $m_1$ and $m_2$ and/or constant
restitution coefficients which depend on the nature of the colliding
grains but not on their velocities, i.e. $r_{11}, r_{22}$ and
$r_{12}=r_{21}$.

The mixture evolves according to the following set of stochastic
equations:

\begin{equation}
m_i \frac{d v_i}{d t}=F_i+f_i+\xi_i(t)
\label{motion}
\end{equation}

where the total force acting on particle $i$ is made of three
contributions: the impulsive force, $F_i$, due to mutual collisions,
the velocity dependent force, $f_i=-\Gamma v_i$, due to the friction
of the particles with the surroundings and the stochastic force,
$\xi_i$, due to an external random drive.  Since we are interested in
the rapid granular flow regime, we model the collisions as
instantaneous binary events, similar to those occurring in a hard
sphere system.

The presence of the frictional, velocity dependent term in addition to
the random forcing \cite{Puglio}, not only mimics the presence of
friction of the particles with the container, but also is motivated by
the idea of preventing the energy of a driven elastic system ($\gamma
\to 1$), to increase indefinitely.

Let us observe that in the absence of collisions the velocity changes
are described by the following Ornstein-Uhlenbeck process:

\begin{equation}
m_{\alpha} \partial_t  
v_i(t)=-\Gamma v_i(t)+\xi_i^{\alpha}(t)
\label{driv}
\end{equation}

where the stochastic acceleration term is assumed to have a white
spectrum with zero mean:

\begin{equation}
\langle \xi_i^{\alpha}(t) \rangle=0
\label{aver}
\end{equation}

and variance:

\begin{equation}
\langle \xi_i^{\alpha}(t)\xi_j^{\beta}(t') \rangle=2 D \delta_{\alpha,\beta}
\delta_{i,j}\delta(t-t')
\label{varian}
\end{equation}

By redefining the bath constants $\Gamma_{\alpha}=\frac{\Gamma}{
m_{\alpha}}$ and $D_{\alpha}=\frac{D}{ m_{\alpha}^2}$, it is
straightforward to obtain the probability distributions of the
velocity of each species, $P_{\alpha}(v,t)$.  In fact, the
Fokker-Planck equations associated with the process (\ref{driv}):

\begin{equation}
\partial_t P_{\alpha}(v,t)=
\Gamma_{\alpha}(\partial_v v P_{\alpha}(v,t))+
D_{\alpha}(\partial^2_{v} P_{\alpha}(v,t))
\label{Fokker}
\end{equation}

possess the following stationary distribution functions:

\begin{equation}
P_{\alpha}(v)=\sqrt{\frac{m_{\alpha}}{2 \pi T_b}} e^{-\frac{m_{\alpha}
v^2}{2 T_b}}
\label{equil}
\end{equation}

where $T_b=\frac{D}{\Gamma}$ represents the temperature of the 
heath-bath, that we fix to be the same for the two species \cite{footnote}.

In order to represent the effect of the collisions on the evolution of
the system we assume that the velocities change instantaneously
according to the rules:

\begin{subequations}
\label{collision}
\begin{align}
v_i^{'\alpha} & =  v_i^{\alpha}-
[1+r_{\alpha \beta}]\frac{m_{\beta}}{m_{\alpha}+
m_{\beta}}(v^{\alpha}_i-v^{\beta}_j) \\
v^{'\beta}_j & =  v^{\beta}_j+
[1+r_{\alpha \beta}]\frac{m_{\alpha}}{m_{\alpha}+m_{\beta}}(v^{\alpha}_i-v^{\beta}_j)
\end{align}
\end{subequations}

where the primed quantities are the post-collisional velocities and
the primed are the velocities before the collisions. A finite fraction
of the kinetic energy of each pair is dissipated during a collision.
Between collisions the velocities perform a random walk due to the
action of the heat-bath. The typical time $\tau_c$ between
particle-particle collision is assumed to be large compared to the
time between random kicks. On the other hand the typical time scales
associated with the bath are
$\tau_{b1}=\frac{m1}{\Gamma}=\frac{1}{\Gamma_1}$ and
$\tau_{b2}=\frac{m_2}{\Gamma}=\frac{1}{\Gamma_2}$. When $\Gamma \to 0$
we must also take $D \to 0$ in performing the elastic limit, otherwise
the kinetic energy would diverge asymptotically. In fact in section V
we shall discuss the situation of inelastic particles with vanishing
friction, a case already considered in refs. \cite{BenNaim}
\cite{Cercignani}.

The evolution equations for the probability densities of finding
particles of species $\alpha$ with velocity $v$ at time $t$ for the
system subject both to external forcing and to collisions are simply
obtained by adding the two effects:

\begin{equation}
\partial_t P_{\alpha}(v,t)\!=\Gamma_{\alpha}(\partial_v v P_{\alpha}(v,t))
+D_{\alpha}(\partial^2_{v} P_{\alpha}(v,t))+
\frac{1}{\tau_c} Q_{\alpha}(P_1,P_2) 
\label{sym}
\end{equation}

where the collision integrals $Q_{\alpha}$ consist of a negative loss
term and a positive gain term respectively:

\begin{subequations} \label{prob}
\begin{align}
\begin{split} \label{prob1}
Q_{1}(P_1,P_2) \!&=-\!P_1(v,t)\!+\!\frac{2 p}{1+r_{11}}
\!\int\!\!\!du\, P_1(u,t)P_1\left(\frac{2 v-(1-r_{11})u}{1+r_{11}},t\right) \\
&+\frac{(1-p)}{1+r_{12}}\frac{m_1+m_2}{m_2}
\int\!\!\!du\,P_1(u,t)P_2\left(\frac{\frac{m_1+m_2}{m_2} v-(\frac{m_1}{m_2}-r_{12})u}{1+r_{12}},t\right)
\end{split} \\
\begin{split} \label{prob2}
Q_{2}(P_1,P_2)\!&=-\!P_2(v,t)\!+\!\frac{2 (1- p)}{1+r_{22}}
\!\int\!\!\!du\,  P_2(u,t)P_2\left(\frac{2 v-(1-r_{22})u}{1+r_{22}},t\right) \\
&+\frac{p}{1+r_{12}}\frac{m_1+m_2}{m_1} \int\!\!\!du\, 
P_2(u,t)P_1\left(\frac{\frac{m_1+m_2}{m_1} v-(\frac{m_2}{m_1}-r_{12})u}{1+r_{12}},t\right)
\end{split}
\end{align}
\end{subequations}

where $p=N_1/(N_1+N_2)$.  In writing eqs. (\ref{prob}) we have assumed
that the collisions occur instantaneously and that collisions
involving more than two particles simultaneously can be
disregarded. Moreover, all pairs are allowed to exchange impulse
regardless of their mutual separation. In this sense we are dealing
with a mean field model.  In order to proceed further it is convenient
to take Fourier transforms of eqs. (\ref{prob}) and
employ the method of characteristic functions \cite{Bobylev} defined
as:

\begin{equation}
\hat P_{\alpha}(k,t)=\int_{-\infty}^{\infty} dv e^{ikv} P_{\alpha}(v,t)
\label{character}
\end{equation}

The resulting equations read:  

\begin{subequations} \label{fou}
\begin{align}
\begin{split} \label{fou1}
\partial_t \hat P_1(k,t) =
-D_1 k^2\hat P_1(k,t) &-\Gamma_1 k \partial_k\hat P_1(k,t)
-\frac{1}{\tau_c}[\hat P_1(k,t)- p \hat P_1(\gamma_{11}k,t) 
\hat P_1((1-\gamma_{11})k,t) \\
&-(1-p) \hat P_1(\tilde\gamma_{12}k,t)
\hat P_2(1-\tilde\gamma_{12})k,t)] 
\end{split} \\
\begin{split}  \label{fou2}
\partial_t \hat P_2(k,t)=
-D_2  k^2\hat P_2(k,t) &-\Gamma_2 k \partial_k\hat P_2(k,t)
-\frac{1}{\tau_c}[\hat P_2(k,t)-(1-p) \hat P_2(\gamma_{22}k,t) 
\hat P_2((1-\gamma_{22})k,t) \\
&-p \hat P_2(\tilde\gamma_{21}k,t)
\hat P_1((1-\tilde\gamma_{21})k,t)] 
\end{split}
\end{align}
\end{subequations}

with 

\begin{subequations}
\begin{align}
\gamma_{\alpha \beta}&=\frac{1-r_{\alpha \beta}}{2}\\
\tilde\gamma_{12}&=[1-\frac{2}{1+\zeta}(1-\gamma_{12})] \\
\tilde\gamma_{21}&=[1-\frac{2}{1+\zeta^{-1}}(1-\gamma_{12})]
\end{align}
\end{subequations}

with $\zeta=m_1/m_2$

 From the mathematical point of view the driven case is very
different from the cooling case. As we have seen in the latter case
the tails are originated by the presence of a singular point at the
origin in the equation for the characteristic function
\cite{mixture}. The singularity is due to 
the conspiracy of the constant cooling rate
and of the scaling form of the evolution equation. 
Such a singularity is removed in the
driven case, thus high velocity tails cease to exist and the distribution 
will be much more well behaved.

The mathematical structure of eqs. (\ref{fou}) is particularly simple
and in fact there exists a standard method of solution. It consists in
expanding the Fourier transform $\hat P_{\alpha}(k,t)$ of the
distributions $P_{\alpha}(v,t)$ in a Taylor series around the origin
$k=0$:

\begin{equation}
\hat P_{\alpha}(k,t)=\sum_{n=0}^{\infty} \frac{(i k)^n}{n !} 
\mu^{\alpha}_{n}(t)
\label{Taylor}
\end{equation}

and substituting (\ref{Taylor}) into eqs. (\ref{fou}).
 
Equating like powers of $k$ we obtain a hierarchy of equations for the
$\mu^{\alpha}_{n}(t)$ which can be solved by a straightforward
iterative method.  At this stage one can appreciate the mathematical
convenience of the Maxwell model. In fact, the coefficients of the
Taylor series represent the moments of the velocity distributions

\begin{equation}
\mu^{\alpha}_{n}(t)=\int_{-\infty}^{\infty} dv v^n P_{\alpha}(v,t)
\label{momento}
\end{equation}

Since the evaluation of the moments of a given order requires only the
knowledge of the moments of lower order one can proceed without
excessive difficulty to any desired order.  In practice, we carried on
our calculation up to the eighth moment, assuming that the initial
distributions were even, so that the odd moments vanish. In order to
render the reading of the paper more expeditious we shall report the
equations determining the stationary value of the moments in the
Appendix.


\section{Two Temperature behavior}

In order to determine the granular temperatures we equate the
coefficients of order $k^2$ in eqs. (\ref{fou}) and obtain the
governing equations for the second moments:

\begin{subequations} \label{t}
\begin{align}
\begin{split} \label{t1}
\tau_c \partial_t \mu^{(1)}_{2}=\{p[2 \gamma_{11}(\gamma_{11}-1)]
&-(1-p)(1-\tilde\gamma_{12}^2)-2 \tau_c \Gamma_1 \}\mu^{(1)}_{2} \\
&+(1-p)(1-\tilde\gamma_{12})^2\mu^{(2)}_{2} +2\tau_c D_1
\end{split} \\
\begin{split} \label{t2}
\tau_c \partial_t \mu^{(2)}_{2}=\{(1-p)[2 \gamma_{22}(\gamma_{22}-1)]
&-p(1-\tilde\gamma_{21}^2)-2\tau_c \Gamma_2 \}\mu^{(2)}_{2}  \\
&+p(1-\tilde\gamma_{21})^2\mu^{(2)}_{1} +2\tau_c D_2
\end{split}
\end{align}
\end{subequations}

The r.h.s. of eqs. (\ref{t}) represent the balance between the energy
dissipation due to inelastic collisions and friction and the energy
input due to the bath.  We define the global and the partial granular
temperatures respectively as $T_g=p T_1+(1-p) T_2$ and
$T_{\alpha}=\frac{1}{2} m_{\alpha}<v_{\alpha}^2>$, where the average
is performed over the noise $\xi_i$.  Since the energy dissipation and
the energy supply mechanisms compete, the system under the influence
of a stochastic white noise driving achieves asymptotically a
statistical steady state. Notice that eqs. (\ref{t}) feature only the
second moments of the velocity distributions, so that the solution is
straightforward. Such a state of affairs should be contrasted with the
analogue problem of determining the partial temperatures in Boltzmann
models \cite{Dufty}.
  
Let us start analyzing the behavior of the Maxwell gas in the one
component limit limit $p \to 1$. The granular temperature approaches
its stationary value $T_1$ exponentially:

\begin{equation}
T_1(t)=T_1(0)e^{-\frac{2 t}{ \tau}}+
T_1(\infty)[1-e^{-\frac{2 t}{\tau}}]
\label{time}
\end{equation}

where the constant $\tau$ represents a combination of the 
two characteristic times of the process given by:

\begin{equation}
\frac{1}{\tau}=\frac{\gamma_{11}(1-\gamma_{11})}{\tau_c}+\frac{\Gamma}{m_1}
\label{ctime}
\end{equation}

which shows that $T_b$ is an upper bound to the granular temperature.
We also obtain a simple relation between the temperature of the bath
and the granular temperature $T_1$:

\begin{equation}
{T_1}_{\infty}=m_1 \frac{D_1}{\Gamma_1-
\frac{\gamma_{11}(\gamma_{11}-1)}{\tau_c}}\leq\frac{D}{\Gamma}=T_{b}
\label{singtemp} 
\end{equation}

On the other hand,
when the two components are not identical 
eqs.(\ref{t}) show that the equilibrium macro-state is
specified by two different partial granular temperatures, both
proportional to the heath bath temperature. Hence, the temperature 
ratio is independent from the driving intensity $D$ as
one can see from the formula:

\begin{equation}
\frac{T_1}{T_2}=\frac{1}{\zeta}\frac{(1-p)[2 \gamma_{22}(1-\gamma_{22})]
+p(1-\tilde\gamma_{21}^2)+2 \frac{\tau_c}{\tau_{b1}}+(1-p)\zeta^2(1-\tilde\gamma_{12})^2}
{p[2 \gamma_{11}(1-\gamma_{11})]
+(1-p)(1-\tilde\gamma_{12}^2)+2 \frac{\tau_c}{\tau_{b2}} 
+p\zeta^{-2}(1-\tilde\gamma_{21})^2}  
\label{ratiot}
\end{equation}

\begin{figure}[h]
\centerline {\includegraphics[clip=true,width=10cm,keepaspectratio, angle=0]{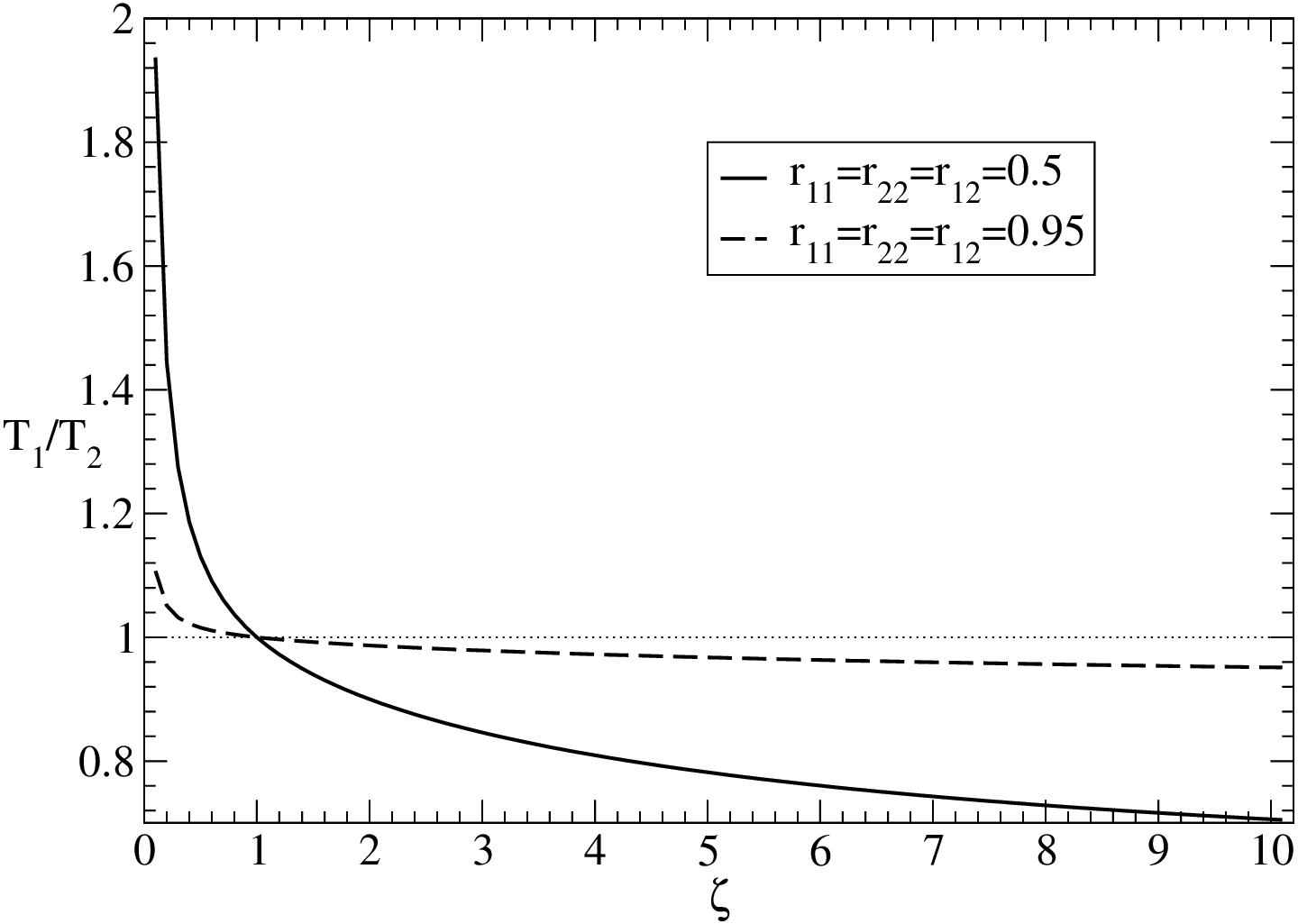}}
\caption
{Temperature ratio as a function of the mass ratio for different
choices of the inelasticity parameter, for $p=0.5$ and
$r_{11}=r_{22}=r_{12}=r=0.8$ (solid line) and $r=0.5$ (dashed-line)}
\label{tempratio}
\end{figure}

Formula in eq. (\ref{ratiot}) illustrates the two temperature behavior
of an inelastic mixture subject to external driving. Notice that the
temperature ratio in the driven case is different from the
corresponding quantity in the cooling undriven case, for the same
model system. In the undriven case we found that the homogeneous
cooling state was characterized by two different exponentially
decreasing temperatures, but whose ratio was constant. However, no
simple relation exists between the ratio relative to the two cases, on
account of the fact that the energy exchanges involved are rather
different.  Thus, in the presence of a heath bath the inelastic
mixture displays the two temperature behavior already reported in the
free cooling case \cite{Dufty,mixture} and in experiments
\cite{Menon}. This feature seems to be a general property of inelastic
systems.  In fig. (\ref{tempratio}) we display the temperature ratio
as a function of the mass ratio $\zeta$ for two different values of
the inelasticity.  Notice that the temperature of the heavier
component is lower than the one of the lighter species.

\begin{figure}[h]
\centerline {\includegraphics[clip=true,width=10cm, keepaspectratio]{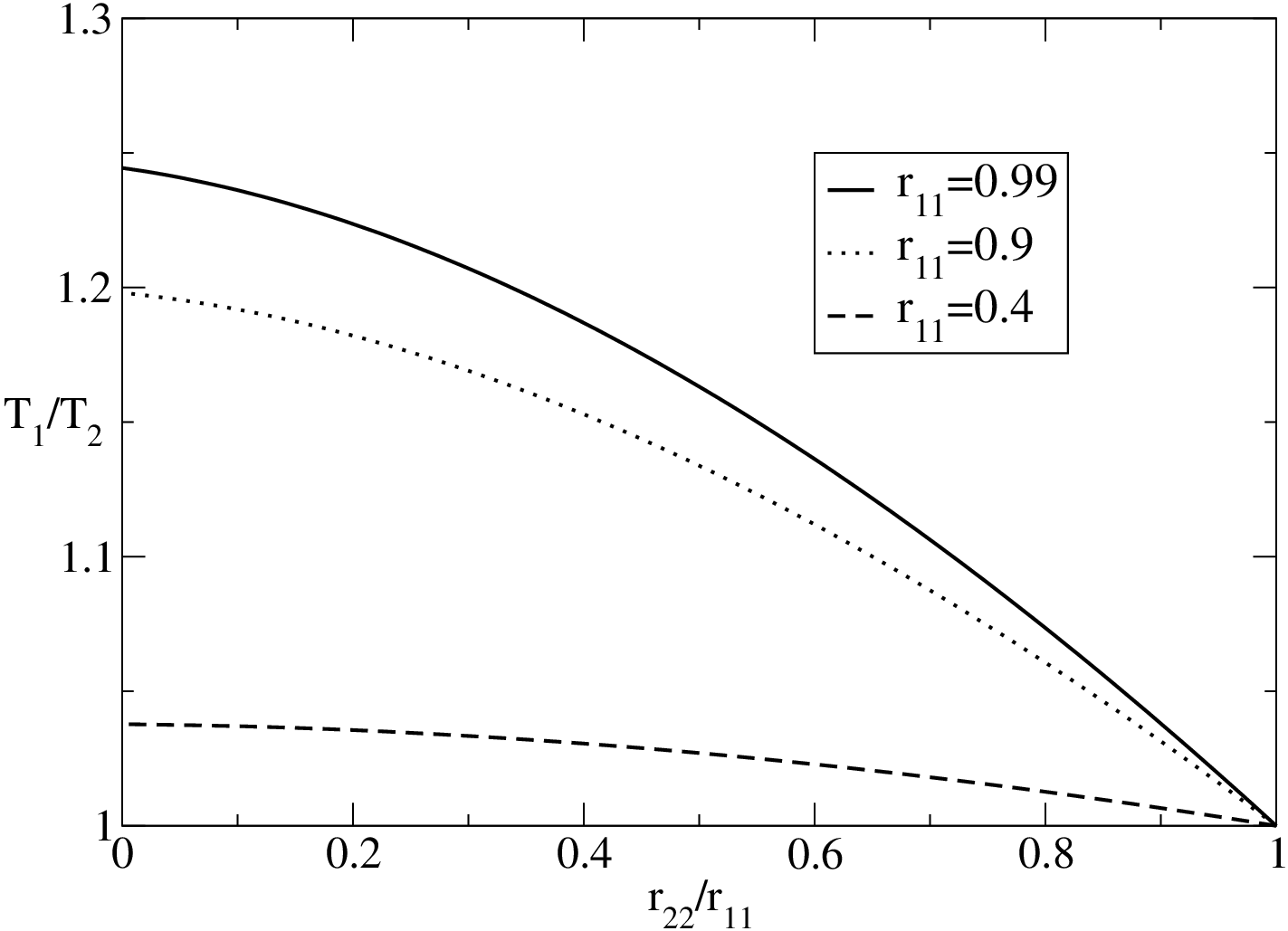}}
\caption
{Temperature ratio as a function of the asymmetry $r_{22}/r_{11}$ for
$p=0.5$, $\zeta=1$ and different 
choices of the inelasticity parameter $r_{11}$,
from top to bottom this is respectively $0.99$, $0.9$, $0.4$ (line)}
\label{ratio_r22_r11}
\end{figure}

In fig. \ref{ratio_r22_r11} the ratio of the temperatures of the two
species is plotted in the case of an asymmetry in the restitution
coefficients parametrized by the form: $r_{22}=r_{11}-x$ with
$r_{12}=(r_{22}+r_{11})/2$, for $p=1/2$, identical masses and three
different values of the coefficient $r_{11}$ as shown in figure.  One
sees that the variation of $\frac{T_1}{T_2}$ is much smaller than the
corresponding variation with respect to the mass asymmetry shown in
fig. \ref{tempratio} in agreement with the experimental observation of
reference \cite{Menon}.  It seems reasonable to conclude that the mass
asymmetry is the larger source of temperature difference between the
two components.



\section{Velocity distribution functions}

An interesting aspect of granular systems concerns the nature of the
single particle velocity distributions. The inelasticity, in fact,
causes marked departures of $P_{\alpha}(v,t)$ from the Gaussian form
which characterizes gases at thermal equilibrium. In undriven gases
these deviations are particularly pronounced and one observes inverse
power law high-velocity tails both in gases of pseudo-Maxwell
molecules \cite{mixture} and in IHS \cite{Ernst}. In the driven case,
i.e. in systems subject to Gaussian white noise forcing (similar to
that represented by eq. (\ref{driv}) with $\Gamma=0$) exponential
tails of the form $exp(-v^{3/2})$ have been predicted theoretically in
inelastic hard-sphere models \cite{Ernst} and tested by direct
simulation Monte Carlo of the Enskog-Boltzmann equation
\cite{Montanero}. Have these non Gaussian tails a counterpart in
Maxwell models?  Ben-Naim and Krapivsky \cite{BenNaim} on the basis of
a re-summation of the moment expansion concluded that the scalar
Maxwell models with vanishing viscosity ($\Gamma=0$) should display
Gaussian-like tails. However, this prediction is in contrast with the
argument, employed by Ernst and van Noije in the case of IHS, which
consists in estimating the tails of the distribution by linearizing
the master equation (\ref{prob}) by neglecting the gain term. This
assumption simplifies the analysis and allows us to reach the
conclusion that the velocity distribution for large $v$ should vanish
as:


\begin{equation}
\lim_{v \to \infty} P(v) \propto \exp{(-v/v_0)}
\label{coda}
\end{equation}

with $v_0^2=D \tau_c$. Clearly such a result is in sharp contrast with
the result of ref. \cite{BenNaim} and seems to indicate that the
Sonine expansion does not reproduce faithfully the high-velocity tails
in the case of Maxwell models with vanishing viscosity. The test of
the limit (\ref{coda}) will be shown in the section 6, where we
illustrate the results of our numerical simulations.

On the other hand, the same kind of asymptotic analysis sketched
above, allows us to conclude that the presence of a viscous damping is
the redeeming feature which renders convergent the Sonine expansion
and the associated Gaussian tails. In fact, with a finite value of
$\Gamma$ the asymptotic solution is of the form:

\begin{equation}
\lim_{v \to \infty} P(v) \propto \exp{(-C v^2)} 
\end{equation}
  
We shall test such a prediction in the remaining part of this section
and study the velocity distributions of the individual species when
$\Gamma \neq 0$ by constructing the solution to the master equation
using the Sonine polynomial expansion method, one of the traditional
approaches to the solution of the Boltzmann equation \cite{burnett}.

We shall also investigate whether the two partial distributions can be
cast into the same functional form upon re-scaling the velocities with
respect to the partial granular thermal velocity, in other words if it
is possible to have a data collapse for the two distributions.

We shall first obtain the steady state values of the first eight
moments as illustrated in the Appendix and then compute the
approximate form of the distribution functions by assuming that these
are Gaussians multiplied by a linear combination of Sonine
polynomials.

Let us begin by writing the following Sonine expansion of the
distribution functions:

\begin{equation}
f_{\alpha}(c)=\frac{1}{\sqrt{\pi}}e^{-c^2}[1+
\sum_{n=1}^{\infty}a^{\alpha}_{n} S_n(c^2)]
\label{fc}
\end{equation}

where $f_{\alpha}(c)$ is the re-scaled distribution defined by: 

\begin{equation}
f_{\alpha}(c)=\sqrt{2\mu^{\alpha}_{2}} P_{\alpha}(v). 
\end{equation}

and $c^2=v^2/2\mu^{\alpha}_{2}$.  The expansion gives the
distributions in terms of the coefficients $a^{\alpha}_{n}$ of the
Sonine polynomials $S_n(c^2)$. In practice, one approximates the
series (\ref{fc}) with a finite number of terms. Since the leading
term is the Maxwellian, the closer the system to the elastic limit,
the less term suffice to describe the state. The expression of the
first polynomials is:

\begin{subequations}
\begin{align}
S_0(c^2)&=1 \\
S_1(c^2)&=\frac{1}{2}-c^2 \\
S_2(c^2)&=\frac{3}{8}-\frac{3}{2}c^2+\frac{1}{2}c^4 \\
S_3(c^2)&=\frac{5}{16}-\frac{15}{8}c^2+\frac{5}{4}c^4-\frac{1}{6}c^6 \\
S_4(c^2)&=\frac{35}{128}-\frac{35}{16}c^2+\frac{35}{16}c^4 
-\frac{7}{12}c^6+\frac{1}{24}c^8
\end{align}
\end{subequations}

In order to obtain the first $m$ values $a^{\alpha}_{m}$, we need to
compute the re-scaled moments $\langle c^n \rangle_{\alpha}$ of the distribution
functions up to order $2m$. These moments are evaluated in Appendix by
means of a straightforward iterative method. At the end, knowing the
re-scaled moments, one obtains the following relation for the
coefficients:

\begin{equation}
a^{\alpha}_{n}=\frac{\langle S_n(c^2) \rangle_{\alpha}}{{\cal N}_n}
\label{an}
\end{equation}

Eq. (\ref{an}) can be proved by imposing the consistency condition:

\begin{equation}
\langle c^n \rangle_{\alpha}=\int_{-\infty}^{\infty}dc c^n f_{\alpha}(c)
\label{match}
\end{equation}
in conjunction with the orthogonality property of the Sonine
polynomials:

\begin{equation}
\int_{-\infty}^{\infty} \frac{1}{\sqrt{\pi}}e^{-c^2} 
S_n(c^2)S_m(c^2)={\cal N}_n \delta_{m,n}
\label{ortho}
\end{equation}

where ${\cal N}_n$ is a normalization constant. Notice that in order
to obtain our results we have not assumed weak inelasticity, therefore
these hold for any value of the restitution coefficients.

The coefficients $a^{\alpha}_{m}$ up to the fourth order in terms of
the re-scaled moments read:

\begin{subequations}
\begin{align} 
a^{\alpha}_1&=0 \label{a1} \\
a^{\alpha}_2&=[1-4\langle c^2\rangle_{\alpha}+\frac{4}{3}\langle c^4\rangle_{\alpha}] \label{a2} \\
a^{\alpha}_3&=[1-6 \langle c^2\rangle_{\alpha}+4 \langle c^4\rangle_{\alpha} - \frac{8}{15}\langle c^6\rangle_{\alpha}] \label{a3}\\
a^{\alpha}_4&=[1-8 \langle c^2\rangle_{\alpha}+8 \langle c^4\rangle_{\alpha} - 
\frac{32}{15}\langle c^6\rangle_{\alpha} +\frac{16}{105}\langle c^8\rangle_{\alpha} ] \label{a4} \\
\end{align}
\end{subequations}

\begin{figure}[h]
\centerline {\includegraphics[clip=true,width=10cm, keepaspectratio,angle=0]{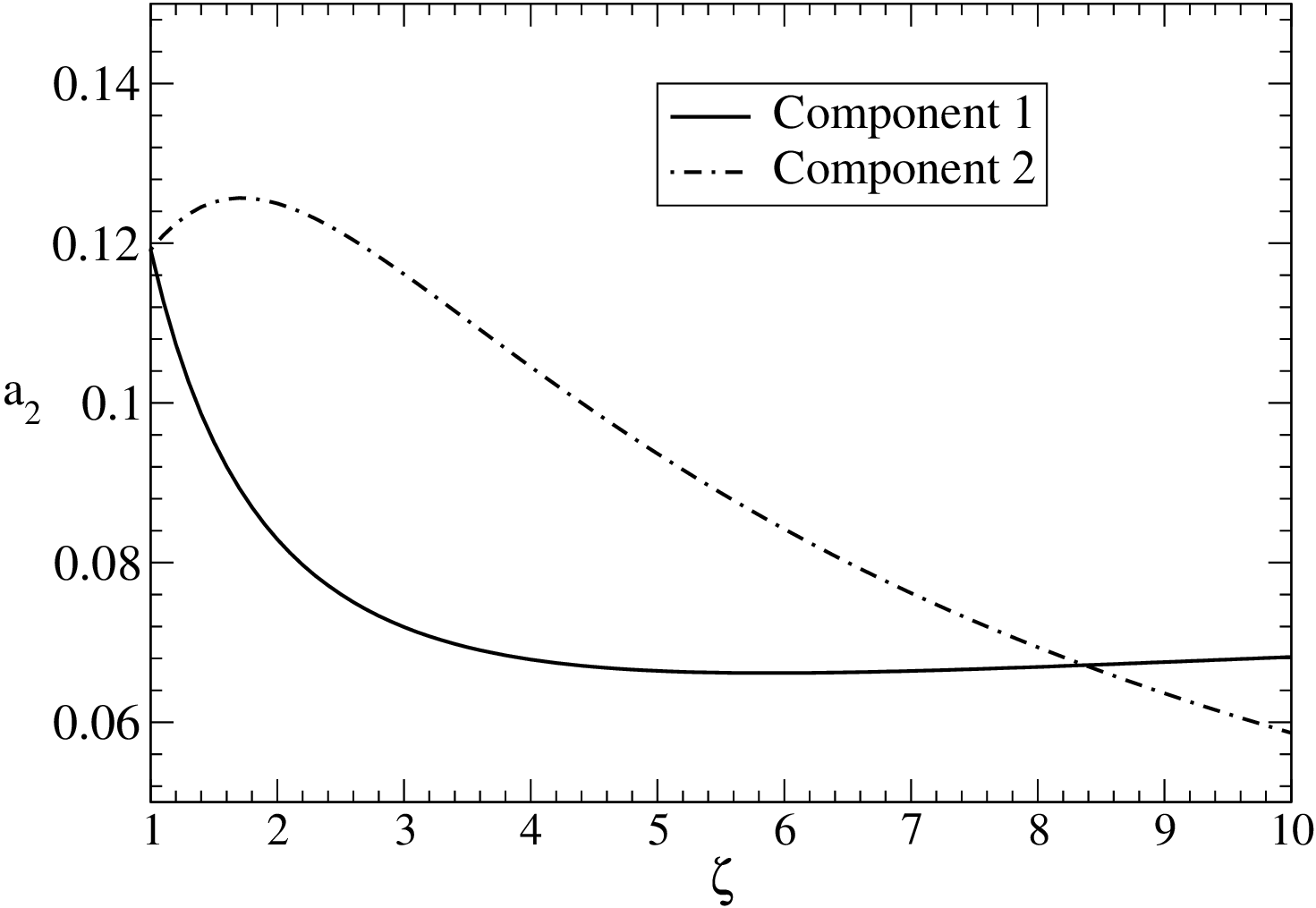}}
\caption
{Second Coefficient of the Sonine expansion $a_2$, for each component
as a function of the mass ratio $\zeta$, $m_1=1$ and for $p=0.5$ and
$r_{11}=r_{22}=r_{12}=0.5$. We have also fixed $\frac{1}{\Gamma}\equiv \tau_{b1}=200$ and
$\tau_c=25$. }
\label{figsonine2}
\end{figure}

\begin{figure}[h]
\centerline {\includegraphics[clip=true,width=10cm, keepaspectratio,angle=0]{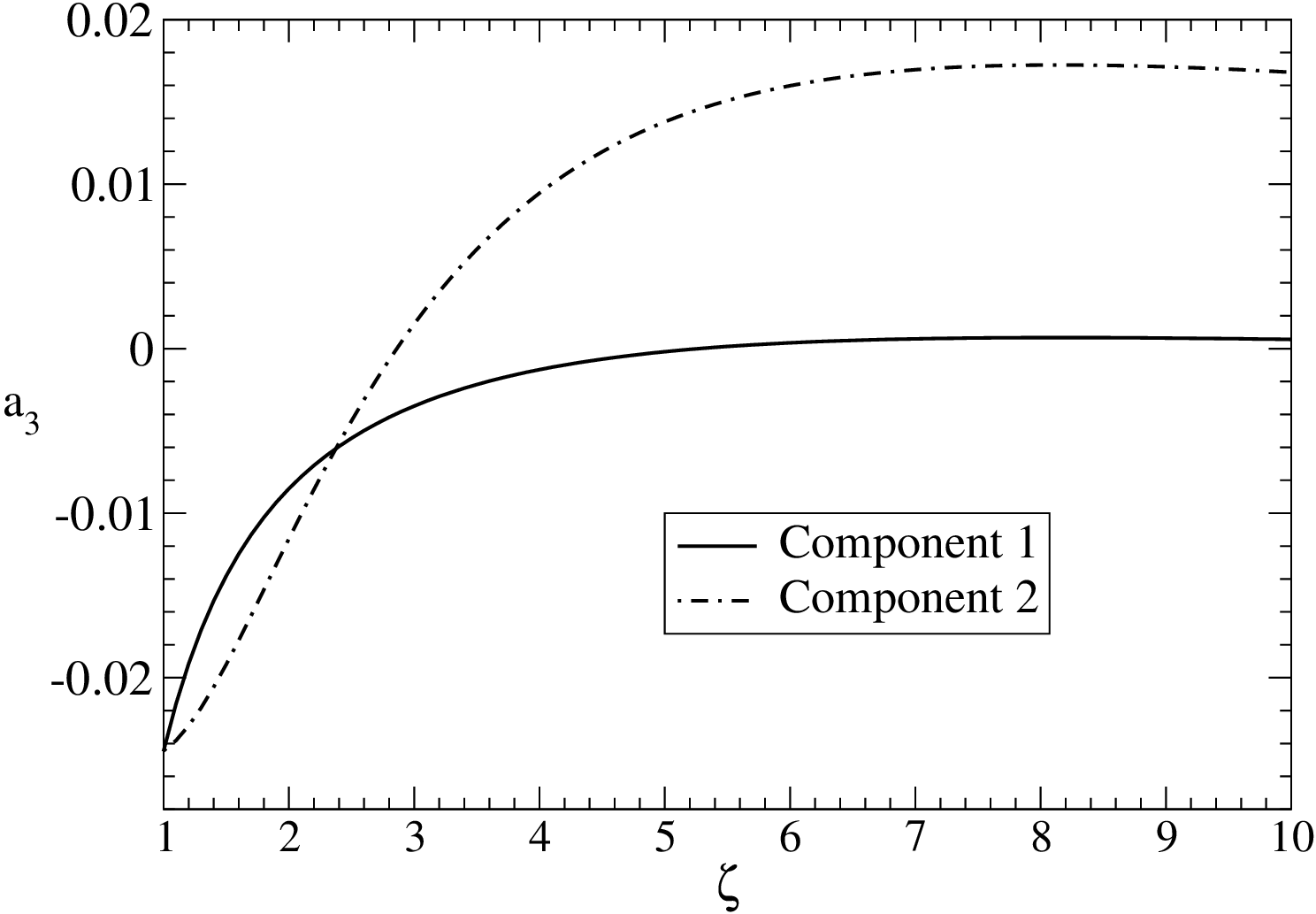}}
\caption
{Third Coefficient of the Sonine expansion $a_3$, for each component
as a function of the mass ratio $\zeta$. The remaining parameters as in
fig. \ref{figsonine2}}
\label{figsonine3}
\end{figure}

\begin{figure}[h]
\centerline {\includegraphics[clip=true,width=10cm, keepaspectratio,angle=0]{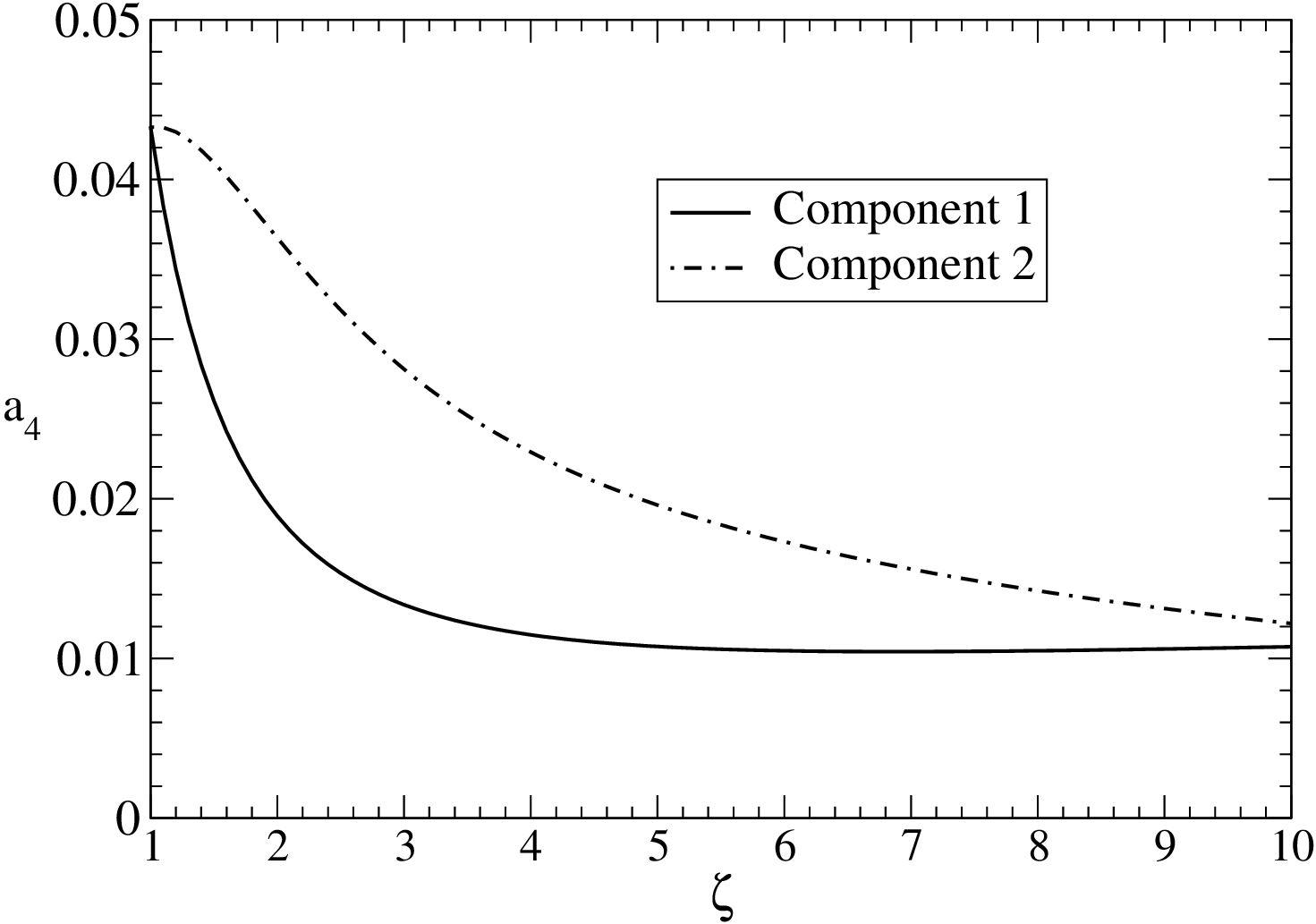}}
\caption
{Fourth Coefficient of the Sonine expansion $a_4$, for each component
as a function of the mass ratio $\zeta$. The remaining parameters as in
fig. \ref{figsonine2}}
\label{figsonine4}
\end{figure}

In fig. (\ref{figsonine2}-\ref{figsonine4}) we display the
behavior of the Sonine coefficients for both components
in the case of equal restitution coefficients $r=0.5$ as
a function of the mass ratio.

\begin{figure}[h]
\centerline {\includegraphics[clip=true,width=10cm, keepaspectratio,angle=0]{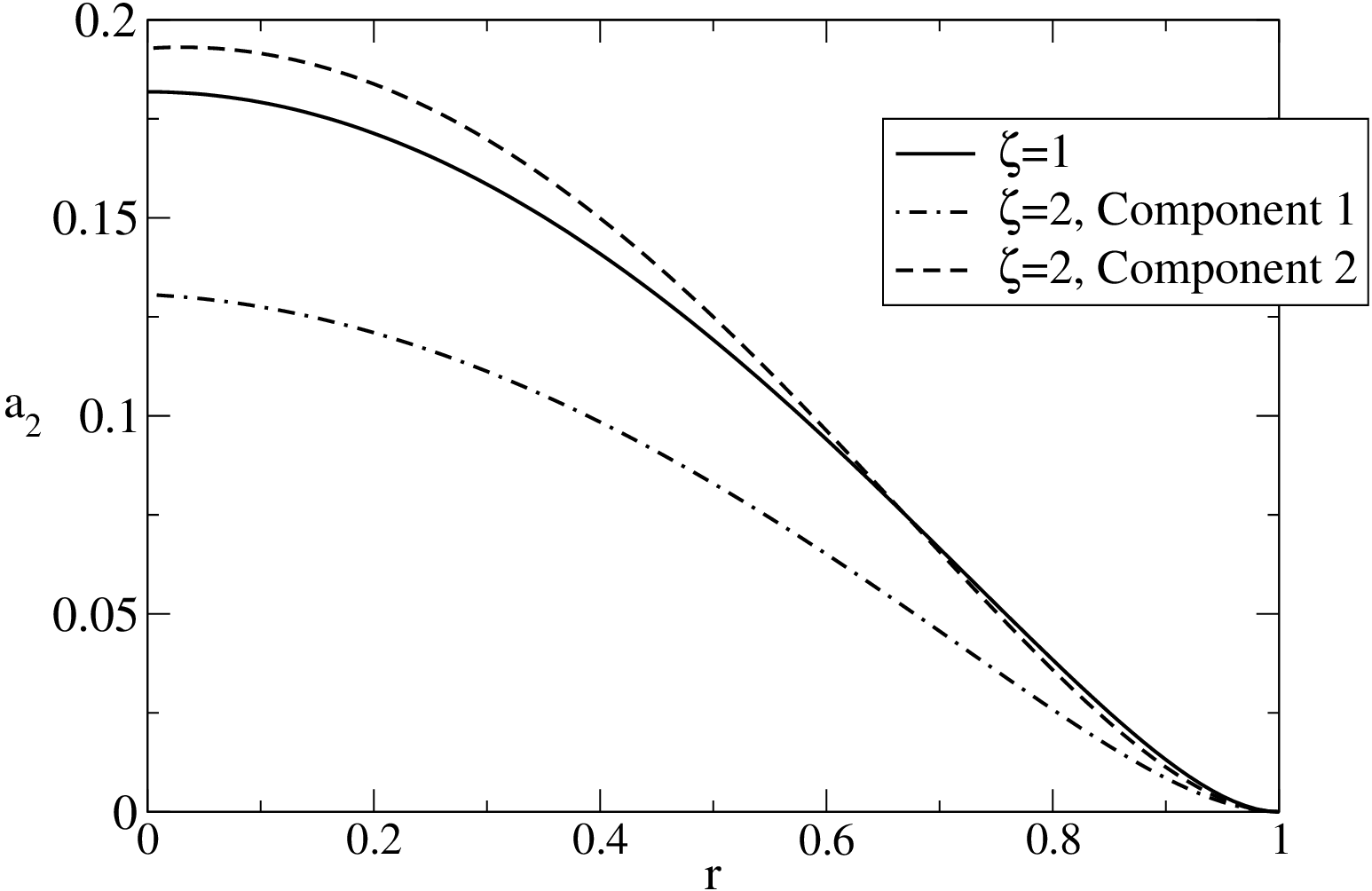}}
\caption
{Second Coefficient of the Sonine expansion $a_2$, for each component
as a function of the inelasticity $r=r_{11}=r_{22}=r_{12}$, for
$p=0.5$ and $m_1=1$ and $\zeta=2$ and $\zeta=1$. The remaining parameters
as in fig. \ref{figsonine2}.}
\label{fsonine2_r}
\end{figure}

\begin{figure}[h]
\centerline {\includegraphics[clip=true,width=10cm, keepaspectratio,angle=0]{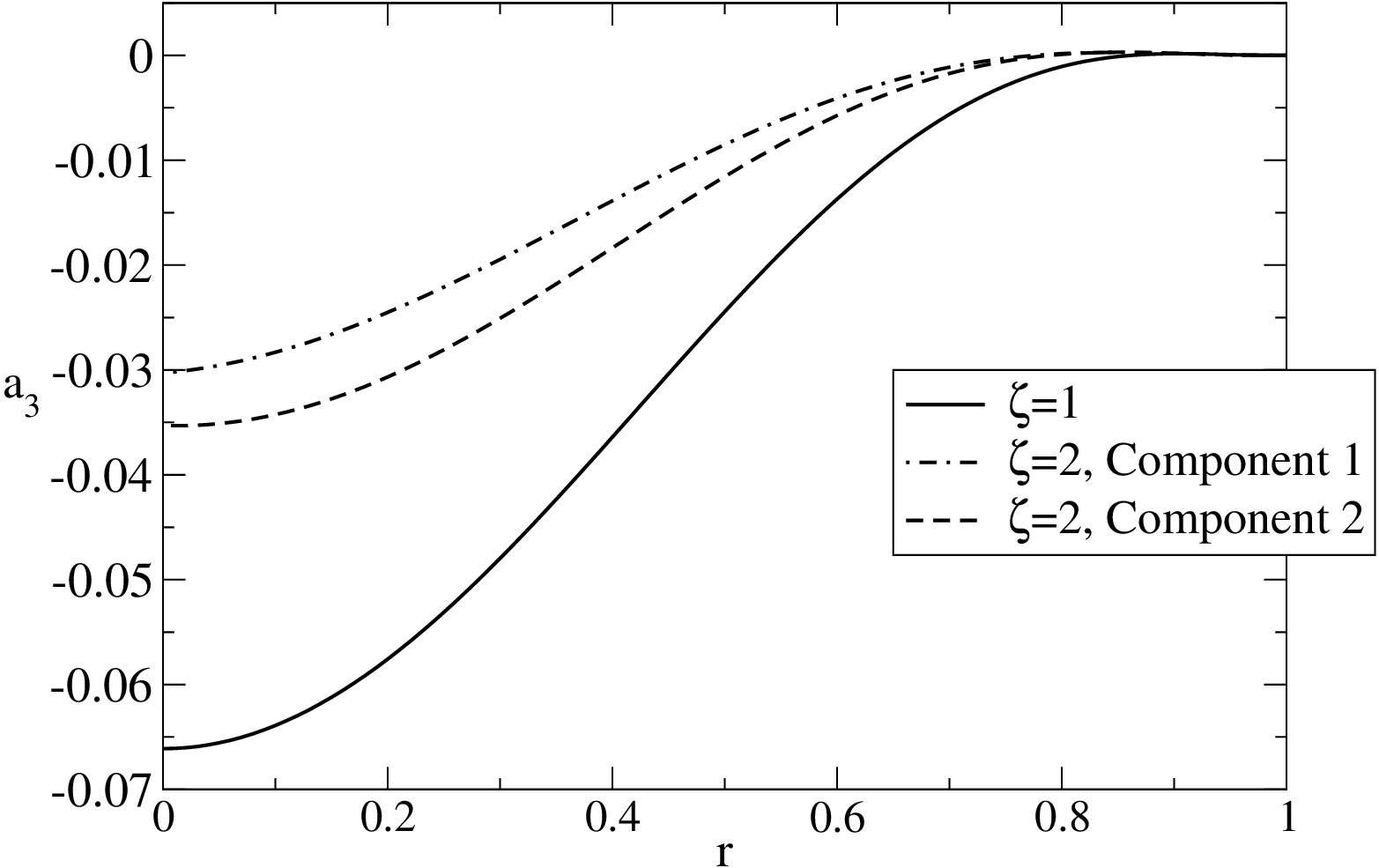}}
\caption
{Third Coefficient of the Sonine expansion $a_3$, for each component
as a function of the inelasticity $r=r_{11}=r_{22}=r_{12}$, for
$p=0.5$ and $m_1=1$ and $\zeta=2$ and $\zeta=1$.The remaining parameters
as in fig. \ref{figsonine2}.} 
\label{fsonine3_r}
\end{figure}

\begin{figure}[h]
\centerline {\includegraphics[clip=true,width=10cm, keepaspectratio,angle=0]{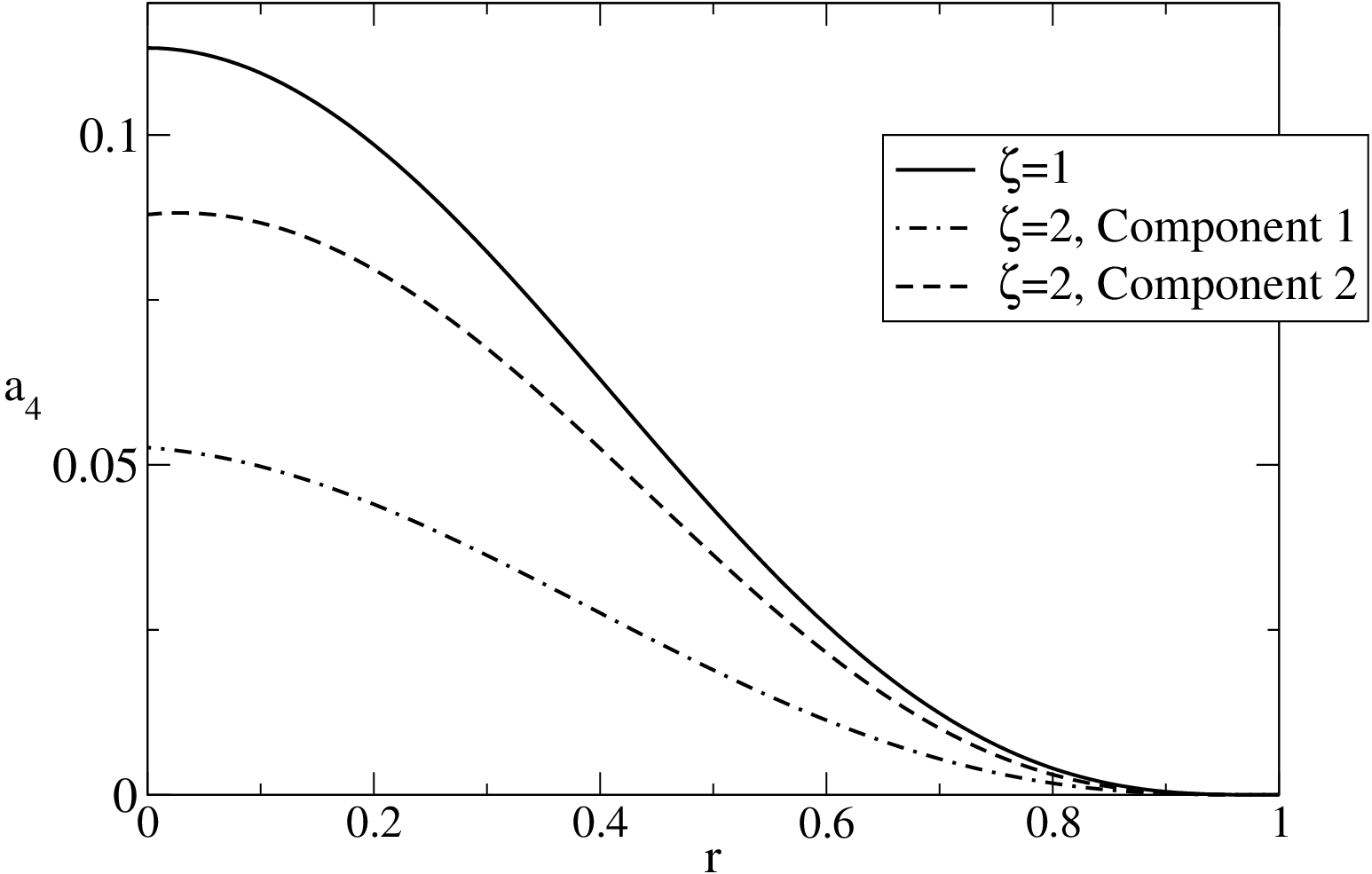}}
\caption
{Fourth Coefficient of the Sonine expansion $a_4$, for each component
as a function of the inelasticity $r=r_{11}=r_{22}=r_{12}$, for
$p=0.5$ and $m_1=1$ and $\zeta=2$ and $\zeta=1$.
The remaining parameters
as in fig. \ref{figsonine2}. }
\label{fsonine4_r}
\end{figure}

In figs. (\ref{fsonine2_r}-\ref{fsonine4_r}) we illustrate the 
variation of the Sonine coefficients for the two components
as a function of the inelasticity for two different values of the
mass ratio: $\zeta=1$ and $\zeta=2$. Notice that the coefficients
are monotonic functions of the inelasticity as already noticed
in pure systems \cite{Ernst}.

\begin{figure}[h]
\centerline {\includegraphics[clip=true,width=10cm, keepaspectratio,angle=0]{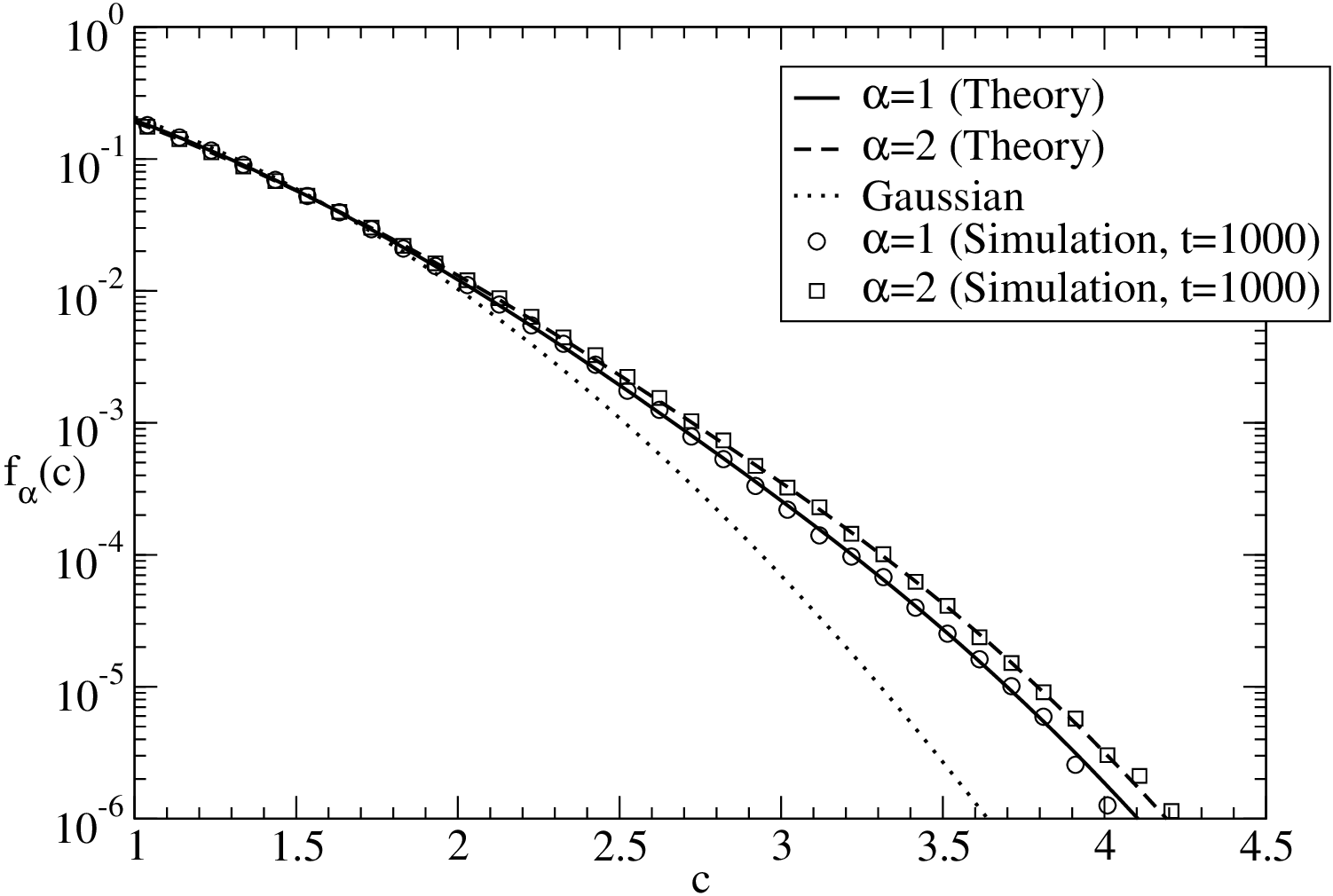}}
\caption
{Rescaled distribution functions for the two components for a model
system characterized by $m_1=1$, $m_2=0.5$ and equal inelasticities
$r=0.5$.$T_b=1$ and the remaining parameters as in
fig. \ref{figsonine2}}
\label{figdistrib}
\end{figure}

In fig.(\ref{figdistrib}) we show the distribution functions for the
heated system with non vanishing viscosity.  The tails become fatter
with increasing order of the approximation, i.e.  the high energy
tails are overpopulated. Moreover, one sees that when $p=1/2$ and the
restitution coefficients are all equal the species with the larger
tails is the lighter.  On the other hand, for a system with the same
masses, but different restitution coefficients, the more elastic
species displays the larger tails.  We also show the numerical data
obtained by simulating the the dynamics. The agreement is quite
satisfactory and validates the approximation method employed.

On the other hand, it is also evident, that the two distributions fail
to collapse one over the other after the rescaling of the
velocities. This fact is consistent with the different values assumed
by the coefficients $a^1_n$ and $a^2_n$. However, this effect is
rather small and can be appreciated only by studying the high velocity
region of the distribution functions.


\section{Numerical Simulations}

To investigate the validity of the previous results and in particular
to test the convergence of the Sonine expansion 
in different situations we shall present in this
section numerical results obtained by simulating an ensemble of $N$
particles subject to a Gaussian forcing, viscous friction and
inelastic collisions.

The scheme consists of the following ingredients:

\renewcommand{\labelenumi}{\roman{enumi}}
\begin{enumerate}

\item
time is discretized, i.e. $t=n\cdot dt$ 

\item
update all the velocities to simulate the
random forcing and the viscous damping:

\begin{equation}
v_i^{\alpha}(t+dt)=
v_i^{\alpha}(t)e^{-\frac{dt}{\tau_b}}+\sqrt{T_b(1-e^{-\frac{2dt}{\tau_b}})}W(t)
\label{discre}
\end{equation}

where $W(t)$ is a normally distributed deviate with zero mean and unit
variance.

\item
Choose randomly $N\frac{dt}{2\tau_c}$ pairs of velocities and update
each of them with the collision rule (\ref{collision}). In this way a
mean collision time $\tau_c$ per particle is guaranteed.

\item 
Change the time counter $n$ and restart from ii.

\end{enumerate}

In other words, at every step each particle experiences a Gaussian
kick thus receiving energy from the bath, whereas it dissipates energy
by collision and by damping.  For example, by choosing $dt=1$,
$m_1/\Gamma=\tau_{b1}=200$ and $\tau_c=25$, we obtain that each
particle in the average experiences $25$ Gaussian kicks between two
successive collisions and that the resulting average kinetic energy is
stationary. In order to compare our numerical simulations with the
theoretical predictions we fixed the temperature of the bath to be
$T_b=1$, i.e. chosen $D=\Gamma$. The results of such simulations are
presented in fig. \ref{figdistrib} and show a very good agreement
between the theory and the simulation.

On the contrary, the agreement between the Sonine expansion and the
simulation is not completely satisfactory when we consider a system
subject to a white noise acceleration, but without viscous friction, a
driving proposed by some authors \cite{Peng}, \cite{Moon}. This can be
considered as the limit $\Gamma \to 0$, $T_b \to \infty$, keeping
constant $D=T_b \Gamma$, in the model defined by eqs. \eqref{sym}: note
that in this case the elastic limit $r \to 1$ cannot be performed
without taking also the limit $D \to 0$ as discussed at the beginning,
in order to avoid a divergence of kinetic energy.  For the sake of
simplicity, we simulated a one component system ($p=1$ and $m=1$) with
vanishing viscosity $\Gamma=0$, but $D=0.0008$, $\tau_c=250$ and
$r=0.5$.  Such a choice yields a granular temperature $T_g=16/15$, as
predicted by our formula (\ref{singtemp}). Notice that in this case
the heath bath temperature diverges and the gas does not have a proper
elastic limit, since all moments diverge when $\gamma
\to 0$. We observed that the tails of the velocity distribution
function are strongly non Gaussian. These decay as a simple
exponential as predicted by our simple analysis of the previous
section. In fig. (\ref{figdistrib}) we report our simulation results
against the Sonine approximation.  We observe that the theoretical
estimate, in spite of incorporating the exact values of the first
eight moments, deviates from the numerical data in the large velocity
region.  In particular the Sonine expansion can only give Gaussian
tails, whereas the simulation indicates a slower exponential
decay. The reason for such a discrepancy is to be ascribed to the slow
convergence of the expansion when $\Gamma=0$.


\begin{figure}[h]
\centerline {\includegraphics[clip=true,width=10cm, keepaspectratio,angle=0]{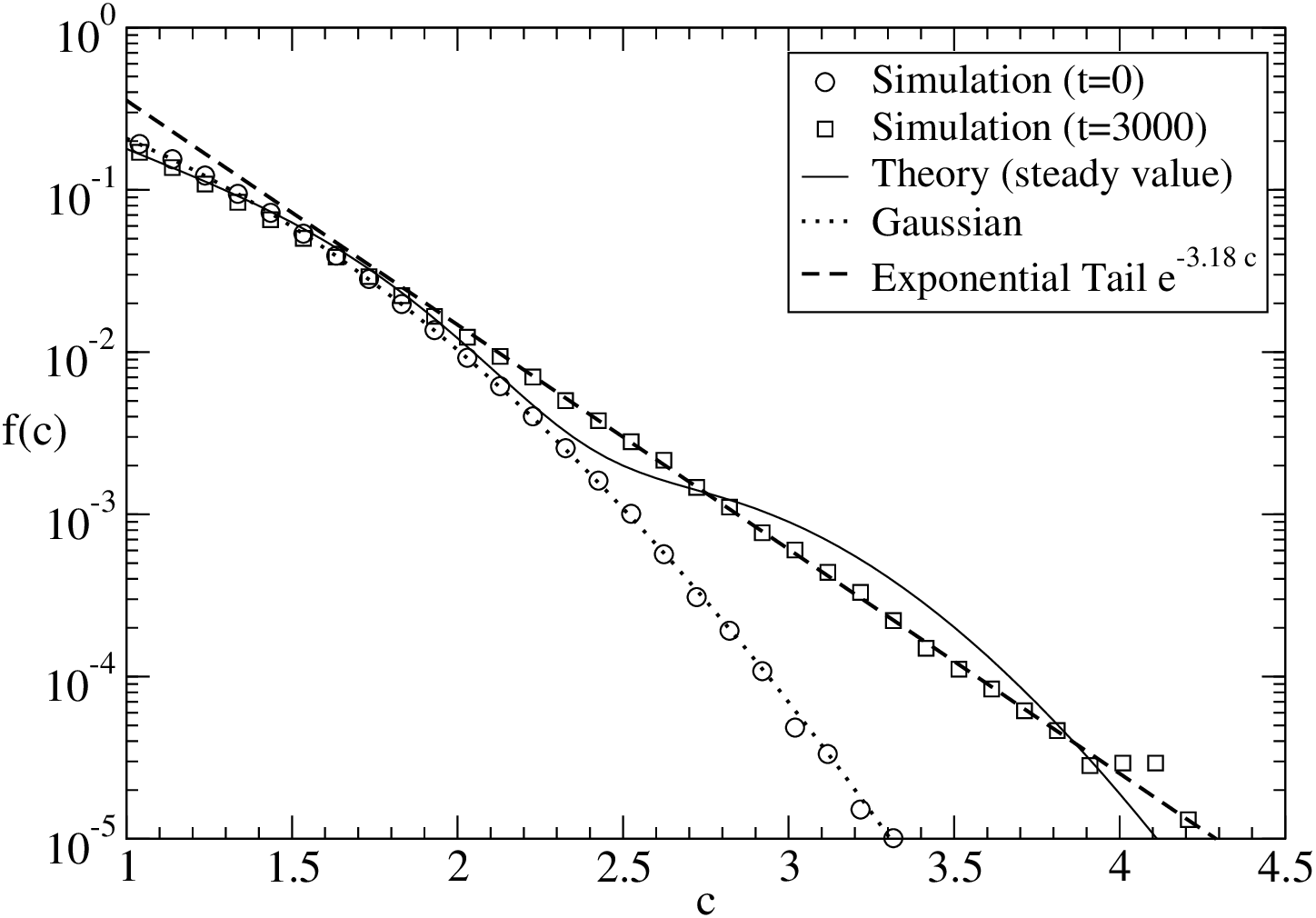}}
\caption
{Rescaled distribution functions for a one component system ($p=1$)
with vanishing viscosity ($\Gamma=0$), $D=0.0008$
, $r=0.5$ and $\tau_c=250$. We show the
initial Gaussian distribution ($t=0$) and the asymptotic 
($t=3000$) stationary
distribution. Notice the presence of high velocity tails. For the sake
of comparison we report the theoretical estimate of the distribution
obtained by means of the Sonine expansion.}
\label{figdistrib2}
\end{figure}


\section{Discussion and Conclusions}

To summarize we have studied the behavior of a model, which perhaps
represent the simplest description of a
driven inelastic gas mixture, namely
an assembly of two types of scalar pseudo-Maxwell molecules subject
to a stochastic forcing. We have obtained 
the velocity distributions 
for arbitrary values of the inelasticity, of the composition and of
the masses by solving the associated Boltzmann equation
by means of a controlled approximation, the moment expansion.
The distributions were obtained by computing exactly the
moments up to the eighth order and then imposing that the corrections
to the Maxwell distribution stemming from the inelasticity are
Gaussians multiplied by a linear combination of Sonine polynomials
with amplitudes determined self-consistently.

The model predicts a steady two temperature behavior which is in
qualitative agreement with existing experimental results. The granular
temperatures can be obtained by very simple algebraic manipulations
for arbitrary values of the control parameters.

By numerical simulations we demonstrated that
the velocity distributions are well described by our series
representation in the case of systems in contact with a bath
at finite temperature $T_b$, whilst the series expansion
breaks down in the case of systems in contact with
bath at infinite temperature, i.e. with zero viscosity.

What can be learned from such a simple model of granular mixture? 
Besides obtaining a global picture of the behavior of the system with
a minimal numerical effort both in the cooling and in the driven case,
the model displays the novel feature of two different distribution
functions, which remain different even after rescaling by the
associated partial granular temperatures.  Of course the detailed form
of the probability velocity distributions are strictly model
dependent, i.e. depend on the assumption of a constant collision rate
inherent in Maxwell models.  Finally, the vectorial character of the
velocities could be included at the cost of a moderate additional
effort. A more interesting and difficult problem would be that of
including in the mixture case a collision frequency proportional to an
appropriate function of the kinetic granular temperatures,
generalizing the work of Cercignani \cite{Cercignani}.


Finally we might ask the general question of the meaning of granular
temperature. Our findings seem to indicate that it is still the main
statistical indicator of the model granular system we studied.
However, with respect to the temperature of a perfectly elastic system
it fails to satisfy a very basic requirement which is known as the
zeroth principle of thermodynamics.

\begin{acknowledgments}
This work was supported by Ministero dell'Istruzione,
dell'Universit\`a e della Ricerca, Cofin 2001 Prot. 2001023848.
\end{acknowledgments}


\appendix*
\section{}

In the present Appendix we sketch the derivation of the various
moments of the distribution functions.  By equating the equal powers
of $k$ in the master equation (\ref{fou}) we obtain a set linear of
coupled equations for the moments. The method of solution is
iterative, because the higher moments depend on the lower
moments. Thus, for instance, to evaluate the lowest order moments of
the distribution functions we must solve the following equations for
the steady state value of the fourth moments:

\begin{subequations}
\begin{align} 
(d^{(4)}_{11}-4 \Gamma_1\tau_c) \mu^{(1)}_{4}+ d^{(4)}_{12} \mu^{(2)}_{4}
+a_{11}(\mu^{(1)}_{2})^2+a_{12} \mu^{(1)}_{2}\mu^{(2)}_{2}
+12 D_1\tau_c \mu^{(1)}_2&=0 \label{qua1} \\
(d^{(4)}_{22}-4 \Gamma_2\tau_c) \mu^{(2)}_{4}+ d^{(4)}_{21} \mu^{(1)}_{4}
+a_{22}(\mu^{(2)}_{2})^2+a_{21} \mu^{(1)}_{2}\mu^{(2)}_{2}+
12 D_2\tau_c \mu^{(2)}_2&=0  \label{qua2}
\end{align}
\end{subequations}

In turn, the sixth moments are obtained by solving:

\begin{subequations}
\begin{align}
(d^{(6)}_{11}-6 \Gamma_1\tau_c) \mu^{(1)}_{6}+ d^{(6)}_{12} \mu^{(2)}_{6}
+b_{11}\mu^{(1)}_{2}\mu^{(1)}_{4}   
+b_{12} \mu^{(1)}_{2}\mu^{(2)}_{4}+b_{12}' \mu^{(2)}_{2}\mu^{(1)}_{4}
+30 D_1\tau_c \mu^{(1)}_4 &=0  \label{ses1} \\
(d^{(6)}_{22}-6 \Gamma_2\tau_c) \mu^{(2)}_{6}+ d^{(6)}_{21} \mu^{(1)}_{6}
+b_{22}\mu^{(2)}_{2}\mu^{(2)}_{4}   
+b_{21} \mu^{(1)}_{2}\mu^{(2)}_{4}+b_{21}' \mu^{(2)}_{2}\mu^{(1)}_{4}
+30 D_2 \tau_c \mu^{(2)}_4 &=0 \label{ses2}
\end{align}
\end{subequations}

Finally the eight moments are the solutions of:

\begin{subequations}
\begin{align}
\begin{split} \label{ott1}
(d^{(8)}_{11}-8 \Gamma_1\tau_c) \mu^{(1)}_{8} &+ d^{(8)}_{12} \mu^{(2)}_{8}
+c_{11}\mu^{(1)}_{2}\mu^{(1)}_{6} +c_{11}'(\mu^{(1)}_{4})^2 \\ 
&+c_{12} \mu^{(1)}_{2}\mu^{(2)}_{6}+c_{12}'\mu^{(1)}_{4}\mu^{(2)}_{4}+
c_{12}'' \mu^{(1)}_{6}\mu^{(2)}_{2} +56 D_1\tau_c \mu^{(1)}_6=0
\end{split} \\
\begin{split} \label{ott2}
(d^{(8)}_{22}-8 \Gamma_2\tau_c) \mu^{(2)}_{8} &+ d^{(8)}_{21} \mu^{(1)}_{8}
+c_{22}\mu^{(2)}_{2}\mu^{(2)}_{6} +c_{22}'(\mu^{(2)}_{4})^2  \\
&+c_{21} \mu^{(2)}_{2}\mu^{(1)}_{6}+c_{21}' \mu^{(2)}_{4}\mu^{(1)}_{4}
+c_{21}'' \mu^{(2)}_{6}\mu^{(1)}_{2} +56 D_2 \tau_c \mu^{(2)}_6=0
\end{split}
\end{align}
\end{subequations}

where the general form of the coefficients $d_{ij}$ is given by:

\begin{subequations}
\begin{align}
d^{(n)}_{11} &=-1+p[\gamma_{11}^n+(1-\gamma_{11})^n]
+(1-p)[\tilde\gamma_{12}]^n \\
d^{(n)}_{12} &=(1-p)[(1-\tilde\gamma_{12})]^n \\
d^{(n)}_{22} &=-1+(1-p)[\gamma_{22}^n+(1-\gamma_{22})^n]
+p[\tilde\gamma_{21}]^n \\
d^{(n)}_{21} &=p [(1-\tilde\gamma_{21})]^n
\end{align}
\end{subequations}

and the coefficients $a_{ij}$ are given by:

\begin{subequations}
\begin{align}
a_{11}&=6 p [\gamma_{11}(1-\gamma_{11})]^2 \\
a_{22}&=6 (1-p) [\gamma_{22}(1-\gamma_{22})]^2 \\
a_{12}&=6(1-p) [(1-\tilde\gamma_{12})]^2 [\tilde\gamma_{12})]^2 \\
a_{21}&=6p [(1-\tilde\gamma_{21})]^2 [\tilde\gamma_{21})]^2 
\end{align}
\end{subequations}

Finally, the coefficients $b_{ij}$ are given by:

\begin{subequations}
\begin{align}
b_{11}&=15 p \gamma_{11}^2(1-\gamma_{11})^2 [\gamma_{11}^2+(1-\gamma_{11})^2] \\
b_{22}&=15 (1-p) \gamma_{22}^2(1-\gamma_{22})^2 [\gamma_{22}^2+(1-\gamma_{22})^2] \\
b_{12}&=15(1-p) [(1-\tilde\gamma_{12})]^4 [\tilde\gamma_{12})]^2 \\
b'_{12}&=15(1-p) [(1-\tilde\gamma_{12})]^2 [\tilde\gamma_{12})]^4 \\
b_{21}&=15p [(1-\tilde\gamma_{21})]^2 [\tilde\gamma_{21})]^4 \\
b'_{21}&=15p [(1-\tilde\gamma_{21})]^4 [\tilde\gamma_{21})]^2 
\end{align}
\end{subequations}

and $c_{ij}$ are

\begin{subequations}
\begin{align}
c_{11}&=28 p \gamma_{11}^2(1-\gamma_{11})^2 [\gamma_{11}^4+(1-\gamma_{11})^4] \\
c_{11}'&=70 p \gamma_{11}^4(1-\gamma_{11})^4 \\
c_{12}&=28(1-p) [(1-\tilde\gamma_{12})]^6 [\tilde\gamma_{12})]^2 \\
c_{12}'&=70(1-p) [(1-\tilde\gamma_{12})]^4 [\tilde\gamma_{12})]^4 \\
c_{12}''&=28(1-p) [(1-\tilde\gamma_{12})]^2 [\tilde\gamma_{12})]^6 \\
c_{22}&=28 (1-p) \gamma_{22}^2(1-\gamma_{22})^2 [\gamma_{22}^4+(1-\gamma_{22})^4] \\
c_{22}'&=70 (1-p) \gamma_{22}^4(1-\gamma_{22})^4 \\
c_{21}&=28p [(1-\tilde\gamma_{21})]^6 [\tilde\gamma_{21})]^2 \\
c_{21}'&=70p [(1-\tilde\gamma_{21})]^4 [\tilde\gamma_{21})]^4 \\
c_{21}''&=28p [(1-\tilde\gamma_{21})]^2 [\tilde\gamma_{21})]^6
\end{align}
\end{subequations}

It is useful to consider the behavior of the moments in the one
component case. The major simplicity of the resulting formulae allows
us to obtain explicit expressions:

\begin{subequations}
\begin{align}
\mu_{2}&=\frac{D\tau_c} {\Gamma\tau_c+\gamma(1-\gamma)}   \\
\mu_{4}&=\frac{12 D\tau_c\mu_2+6 \gamma^2(1-\gamma)^2\mu_{2}^2}{4\Gamma\tau_c+1-\gamma^4-(1-\gamma)^4}  \\
\mu_{6}&=\frac{30 D\tau_c\mu_4+15 \gamma^2(1-\gamma)^2(\gamma^2+(1-\gamma)^2\mu_2\mu_{4}}{6\Gamma\tau_c+1-\gamma^6-(1-\gamma)^6}  \\
\mu_{8}&=\frac{56 D\tau_c\mu_6+
28 \gamma^2(1-\gamma)^2(\gamma^4+(1-\gamma)^4)\mu_2\mu_{6}
+70\gamma^4(1-\gamma)^4\mu_{4}^2}
{8\tau_c\Gamma+1-\gamma^8-(1-\gamma)^8} 
\end{align}
\end{subequations}



\begin{thebibliography}{99}
\bibitem{footnote}
Notice that, in order to reduce the number of free parameters, 
we have assumed the constants, $D$ and $\Gamma$, 
which specify the interactions with the bath 
to be species independent. Such a choice implies that in the
elastic limit both components reach the same granular temperature
given by the temperature of the bath .
\end{thebibliography}
\end{document}